\documentclass[12pt]{article}
\usepackage{amssymb}
\renewcommand{\theequation}{\arabic{section}.\arabic{equation}}

\font\oneeight=cmr10 at 18pt

\newcommand{\vTm}{\vphantom{\mbox{\oneeight I}}}

\begin{document}



\def\a{\alpha}
\def\b{\beta}
\def\d{\delta}
\def\e{\epsilon}
\def\g{\gamma}
\def\h{\mathfrak{h}}
\def\k{\kappa}
\def\l{\lambda}
\def\o{\omega}
\def\p{\wp}
\def\r{\rho}
\def\t{\tau}
\def\s{\sigma}
\def\z{\zeta}
\def\x{\xi}
 \def\A{{\cal{A}}}
 \def\B{{\cal{B}}}
 \def\C{{\cal{C}}}
 \def\D{{\cal{D}}}
\def\G{\Gamma}
\def\K{{\cal{K}}}
\def\O{\Omega}
\def\P{{\cal{P}}}
\def\R{\bar{R}}
\def\T{{\cal{T}}}
\def\L{\Lambda}
\def\f{E_{\tau,\eta}(sl_2)}
\def\E{E_{\tau,\eta}(sl_n)}
\def\Zb{\mathbb{Z}}
\def\Cb{\mathbb{C}}

\def\R{\overline{R}}

\def\no{\nonumber}
\def\le{\langle}
\def\re{\rangle}
\def\lt{\left}
\def\rt{\right}

\baselineskip=20pt

\newfont{\elevenmib}{cmmib10 scaled\magstep1}
\newcommand{\preprint}
{   \begin{flushleft}
   \end{flushleft}\vspace{-1.3cm}
   \begin{flushright}\normalsize
   \end{flushright}}
\newcommand{\Title}[1]{{\baselineskip=26pt
   \begin{center} \Large \bf #1 \\ \ \\ \end{center}}}
\newcommand{\Author}{\begin{center}
   \large \bf
Wen-Li Yang$,{}^{a,b}$
 ~ and~Yao-Zhong Zhang ${}^b$\end{center}}
\newcommand{\Address}{\begin{center}
     ${}^a$ Institute of Modern Physics, Northwest University,
     Xian 710069, P.R. China\\
     ${}^b$ Department of Mathematics, University of Queensland, Brisbane, QLD 4072,
     Australia
   \end{center}}
\newcommand{\Accepted}[1]{\begin{center}
   {\large \sf #1}\\ \vspace{1mm}{\small \sf Accepted for Publication}
   \end{center}}

\preprint
\thispagestyle{empty}
\bigskip\bigskip\bigskip

\Title{Multiple reference states  and complete spectrum of the
$\Zb_n$ Belavin model with open boundaries } \Author

\Address
\vspace{1cm}

\begin{abstract}
The multiple reference state structure of the $\Zb_n$ Belavin
model with non-diagonal boundary terms is discovered. It is found
that there exist $n$ reference states, each of them yields a set
of eigenvalues and Bethe Ansatz equations of the transfer matrix.
These $n$ sets of eigenvalues together constitute the complete
spectrum of the model. In the quasi-classic limit, they give the
complete spectrum of the corresponding Gaudin model.

\vspace{1truecm} \noindent {\it PACS:} 75.10.Pq; 04.20.Jb;
05.50.+q

\end{abstract}
\newpage
\section{Introduction}
\label{intro} \setcounter{equation}{0}

Our understanding of quantum phase transitions and critical
phenomena has been greatly enhanced by the study of exactly
solvable models (integrable models) \cite{Bax82}. Such exact
results provide valuable insights into the key theoretical
development of universality classes in areas ranging from modern
condensed physics \cite{Kas98} to string and super-symmetric
Yang-Mills theories \cite{Dol03}. Among solvable models, elliptic
ones stand out as a particularly important class due to the fact
that most other ones can be obtained from them by some
trigonometric or rational limits. In this paper, we focus on the
elliptic $\Zb_n$ Belavin model \cite{Bel81} with integrable
boundary conditions, with the celebrated XYZ spin chain as the
special case of $n=2$.

Two-dimensional integrable models have traditionally been solved
by imposing periodic boundary conditions. For such bulk systems,
the quantum Yang-Baxter equation (QYBE)
\begin{eqnarray}
 R_{12}(u_1-u_2)R_{13}(u_1-u_3)R_{23}(u_2-u_3)=
 R_{23}(u_2-u_3)R_{13}(u_1-u_3)R_{12}(u_1-u_2), \label{QYB}
\end{eqnarray} leads to families of commuting row-to-row
transfer matrix which may be diagonalized  by the quantum inverse
scattering method (QISM) (or algebraic Bethe Ansatz) \cite{Kor93}.

Not all boundary conditions are compatible with integrability in
the bulk. The bulk integrability is only preserved  when one
imposes certain boundary conditions. In \cite{Skl88}, Sklyanin
developed the boundary QISM, which may be used to describe
integrable systems on a finite interval with independent boundary
conditions at each end. This boundary QISM uses a new algebraic
structure, the reflection equation (RE) algebra. The solutions to
the RE and its dual are called boundary K-matrices which in turn
give rise to boundary conditions compatible with the integrability
of the bulk model \cite{Skl88,Mez91,Gho94}.

The boundary QISM has been successfully applied to diagonalize the
double-row transfer matrices of various integrable models with
non-trivial boundary conditions mostly corresponding to the
diagonal K-matrices. The problem of diagonalizing the double-row
transfer matrix for most general {\it non-diagonal\/} K-matrices
by the algebraic Bethe Ansatz has been  long-standing due to the
difficulty of finding suitable reference states (or pseudo-vacuum
states). Recently, much progress has been made for the open XXZ
spin chain. Bethe Ansatz solutions for non-diagonal boundary terms
where the boundary parameters obey some constraints have been
proposed by various approaches\footnote{Solutions with arbitrary
boundary parameters were recently proposed by functional Bethe
Ansatz \cite{Mur06} and q-Onsager algebra \cite{Bas07}. However,
it seems highly non-trivial to rederive these results in the
framework of algebraic Bethe Ansatz .}
\cite{Nep04,Cao03,Doi03,Yan041,Gie04,Mel05,Yan06,Baj06}. It has
been found that in order to obtain the complete spectrum of the
model two sets of Bethe Ansatz equations and consequently two sets
of eigenvalues are {\it needed} \cite{Nep03,Yan06}, in contrast
with the diagonal boundary case \cite{Skl88}. This suggests that
in the framework of algebraic Bethe Ansatz there should exist two
reference states corresponding to the two sets of Bethe Ansatz
equations and eigenvalues. Such multiple reference state structure
was confirmed in our recent work \cite{Yan07}.

However, for models related to higher rank algebras
\cite{Yan04,Gal05,Yan04-4,Yan05,Yan051,Gal07}  only one reference
state, and consequently only one set of eigenvalues and Bethe
Ansatz equations of their transfer matrices, have been constructed
so far. It is natural to expect that there  exist {\it extra}
reference states, giving rise to extra sets of eigenvalues and
associated Bethe Ansatz equations for such models. In this paper
we investigate the multiple reference state structure for an
elliptic model related to $A_{n-1}$ algebra - the $\Zb_n$ Belavin
model with general non-diagonal boundary terms. It is found that
there actually {\it exist} $n$ reference states for such a model.
Each of these reference states yields a set of eigenvalues and
corresponding Bethe Ansatz equations of the transfer matrix. The
$n$ sets of eigenvalues together constitute the complete spectrum
of the model. In the quasi-classic limit, they give the
corresponding spectrum of the associated Gaudin model
\cite{Gau76}.

The paper is organised as follows. In section 2, we briefly review
the $\Zb_n$ Belavin model with integrable open boundary
conditions, which also serves as an introduction to our notion and
basic ingredients. In section 3, we introduce the intertwiner
vectors and the corresponding face-vertex correspondence relations
which will play key roles in transforming the model in ``vertex
picture" to the one in the ``face picture". In section 4, after
finding the $n$ reference states, we use the algebraic Bethe
Ansatz to obtain the corresponding $n$ sets of eigenvalues and the
associated $n$ sets of Bethe Ansatz equations of the transfer
matrix of the model. In section 5, we take the quasi-classic limit
to extract the spectrum of the associated Gaudin operators.
Section 6 is for conclusion. The Appendix provides the definitions
of some elementary functions appeared in section 4 and 5.


\section{$\Zb_n$ Belavin model
with open boundaries}
\label{Zn} \setcounter{equation}{0}

Let us fix a positive integer $n\geq 2$, a complex number $\tau$
such that $Im(\tau)>0$ and a generic complex number $w$. Introduce
the following elliptic functions
\begin{eqnarray}
\theta\lt[
 \begin{array}{c}
 a\\b
 \end{array}\rt](u,\tau)&=&\sum_{m=-\infty}^{\infty}
 exp\lt\{\sqrt{-1}\pi\lt[(m+a)^2\tau+2(m+a)(u+b)\rt]\rt\},\\
 \theta^{(j)}(u)&=&\theta\lt[\begin{array}{c}\frac{1}{2}-\frac{j}{n}\\
 [2pt]\frac{1}{2}
 \end{array}\rt](u,n\tau),\qquad
 \s(u)=\theta\lt[\begin{array}{c}\frac{1}{2}\\[2pt]\frac{1}{2}
 \end{array}\rt](u,\tau),\label{Function}\\
 \zeta(u)&=&\frac{\partial}{\partial u}\lt\{\ln
 \s(u)\rt\}.\label{Z-function}
\end{eqnarray}
Among them the $\s$-function\footnote{Our $\s$-function is the
$\vartheta$-function $\vartheta_1(u)$ \cite{Whi50}. It has the
following relation with the {\it Weierstrassian\/} $\s$-function
denoted  by $\s_w(u)$: $\s_w(u)\propto e^{\eta_1u^2}\s(u)$,
$\eta_1=\pi^2(\frac{1}{6}-4\sum_{n=1}^{\infty}\frac{nq^{2n}}{1-q^{2n}})
$ and $q=e^{\sqrt{-1}\tau}$. Consequently, our $\zeta$-function
(\ref{Z-function}) is different from the {\it Weierstrassian\/}
$\zeta$-function by an additional term $-2\eta_1 u$.}
 satisfies the following
identity:
\begin{eqnarray}
 &&\s(u+x)\s(u-x)\s(v+y)\s(v-y)-\s(u+y)\s(u-y)\s(v+x)\s(v-x)\no\\
 &&\qquad\qquad \qquad =\s(u+v)\s(u-v)\s(x+y)\s(x-y).\no
\end{eqnarray}
Let $g,\,h,$ be
$n\times n$ matrices with the elements
\begin{eqnarray}
 &&h_{ij}=\d_{i+1\,j},~~g_{ij}=\o^i\d_{i\,j},~~{\rm
 with}\quad \o=e^{\frac{2\pi\sqrt{-1}}{n}},~~~i,j\in \Zb_n.\no
\end{eqnarray}
For any $\a=(\a_1,\a_2)$, $\a_1,\,\a_2\in \Zb_n$, one introduces
an $n\times n$ matrix $I_{\a}$ by
\begin{eqnarray}
 &&I_{\a}=I_{(\a_1,\a_2)}=g^{\a_2}h^{\a_1},\no
\end{eqnarray}
and an elliptic function $\s_{\a}(u)$ by
\begin{eqnarray}
 &&\s_{\a}(u)=\theta\lt[\begin{array}{c}\frac{1}{2}+\frac{\a_1}{n}\\[2pt]
 \frac{1}{2}+\frac{\a_2}{n}
 \end{array}\rt](u,\tau),~~{\rm and\/}~\s_{(0,0)}(u)=\s(u).\no
\end{eqnarray}
The $\Zb_n$ Belavin R-matrix  is \cite{Bel81,Ric86}
\begin{eqnarray}
 R^B(u)=\frac{\s(w)}{\s(u+w)}\sum_{\a\in\Zb_n^2}
 \frac{\s_{\a}(u+\frac{w}{n})}
 {n\s_{\a}(\frac{w}{n})}I_{\a}\otimes
 I_{\a}^{-1}.\label{Belavin-R}
\end{eqnarray}
The R-matrix satisfies the QYBE (\ref{QYB}) and the properties
\cite{Ric86},
\begin{eqnarray}
 &&\hspace{-1.5cm}\mbox{
 Unitarity}:\hspace{42.5mm}R^B_{12}(u)R^B_{21}(-u)= {\rm id},\label{Unitarity}\\
 &&\hspace{-1.5cm}\mbox{
 Crossing-unitarity}:\quad (R^B)^{t_2}_{21}(-u-nw)(R^B)_{12}^{t_2}(u)
 = \frac{\s(u)\s(u+nw)}{\s(u+w)\s(u+nw-w)}\,\mbox{id},
 \label{crosing-unitarity}\\
 &&\hspace{-1.5cm}\mbox{ Quasi-classical
 property}:\hspace{22.5mm}\, R^B_{12}(u)|_{w\rightarrow 0}= {\rm
id}.\label{quasi}
\end{eqnarray}
Here $R^B_{21}(u)=P_{12}R^B_{12}(u)P_{12}$ with $P_{12}$ being the
usual permutation operator and $t_i$ denotes the transposition in
the $i$-th space. Here and below we adopt the standard notation:
for any matrix $A\in {\rm End}(\Cb^n)$, $A_j$ is an operator
embedded  in the tensor space $\Cb^n\otimes \Cb^n\otimes\cdots$,
which acts as $A$ on the $j$-th space and as an identity on the
other factor spaces; $R_{ij}(u)$ is an embedding operator of
R-matrix in the tensor space, which acts as an identity on the
factor spaces except for the $i$-th and $j$-th ones. The
quasi-classical properties (\ref{quasi}) of the R-matrix enables
one to introduce the  associated classical $\Zb_n$  r-matrix
$r(u)$ \cite{Hou99}
\begin{eqnarray}
 R^B(u)&=&{\rm id} +w\,r(u)+O(w^2),\qquad\qquad  {\rm
 when}~w\longrightarrow 0,\no\\[4pt]
 r(u)&=&\frac{1-n}{n}\zeta(u)+\hspace{-0.2truecm}
 \sum_{\a\in\Zb_n^2-(0,0)}\frac{\s'(0)\s_{\a}(u)}
 {n\s(u)\s_{\a}(0)}I_{\a}\otimes
 I_{\a}^{-1},~~\s'(0)=\frac{\partial}{\partial
 u}\s(u)|_{u=0}.\label{r-matrix}
\end{eqnarray}
In the above equation, the
elliptic $\zeta$-function is defined in (\ref{Z-function}).

One introduces  the ``row-to-row" monodromy matrix $T(u)$
\cite{Kor93}, which is an $n\times n$ matrix with elements being
operators acting  on $(\Cb^n)^{\otimes N}$  \begin{eqnarray}
T(u)=R^B_{01}(u+z_1)R^B_{02}(u+z_2)\cdots
R^B_{0N}(u+z_N).\label{T-matrix}\end{eqnarray} Here $\{z_i|i=1,\cdots, N\}$
are arbitrary free complex parameters which are usually called
inhomogeneous parameters. With the help of the QYBE (\ref{QYB}),
one can show that $T(u)$ satisfies the so-called ``RLL" relation
\begin{eqnarray}
R^B_{12}(u-v)T_1(u)T_2(v)=T_2(v)T_1(u)R^B_{12}(u-v).\label{Relation1}\end{eqnarray}

An integrable open chain can be constructed as follows
\cite{Skl88}. Let us introduce a pair of K-matrices $K^-(u)$ and
$K^+(u)$. The former satisfies the RE
 \begin{eqnarray}
 &&R^B_{12}(u_1-u_2)K^-_1(u_1)R^B_{21}(u_1+u_2)K^-_2(u_2)\no\\
  &&~~~~~~=
 K^-_2(u_2)R^B_{12}(u_1+u_2)K^-_1(u_1)R^B_{21}(u_1-u_2),\label{RE-V}
\end{eqnarray}
and the latter  satisfies the dual RE
\begin{eqnarray}
 &&R^B_{12}(u_2-u_1)K^+_1(u_1)R^B_{21}(-u_1-u_2-nw)K^+_2(u_2)\no\\
 &&~~~~~~=
 K^+_2(u_2)R^B_{12}(-u_1-u_2-nw)K^+_1(u_1)R^B_{21}(u_2-u_1).
 \label{DRE-V}
\end{eqnarray}
For the models with open boundaries, instead of
the standard ``row-to-row" monodromy matrix $T(u)$
(\ref{T-matrix}), one needs  the
 ``double-row" monodromy matrix $\mathbb{T}(u)$
\begin{eqnarray}
 \mathbb{T}(u)=T(u)K^-(u)T^{-1}(-u).\label{Mon-V-1}
\end{eqnarray}
Using
(\ref{Relation1}) and (\ref{RE-V}), one can prove that
$\mathbb{T}(u)$ satisfies
\begin{eqnarray}
 R^B_{12}(u_1-u_2)\mathbb{T}_1(u_1)R^B_{21}(u_1+u_2)
  \mathbb{T}_2(u_2)=
 \mathbb{T}_2(u_2)R^B_{12}(u_1+u_2)\mathbb{T}_1(u_1)R^B_{21}(u_1-u_2).
 \label{Relation-Re}
\end{eqnarray}
Then the {\it double-row transfer
matrix\/} of  the inhomogeneous $\Zb_n$ Belavin model  with open
boundary is given by
\begin{eqnarray}
\t(u)=tr(K^+(u)\mathbb{T}(u)).\label{trans}
\end{eqnarray}
The commutativity of the transfer matrices
\begin{eqnarray}
 [\t(u),\t(v)]=0,\label{Com-2}
\end{eqnarray}
follows as a consequence of (\ref{QYB})-(\ref{crosing-unitarity})
and (\ref{RE-V})-(\ref{DRE-V}). This ensures the integrability of
the inhomogeneous $\Zb_n$ Belavin model with open boundary.

In this paper, we consider a {\it non-diagonal\/} K-matrix
$K^{-}(u)$ which is a solution to the RE (\ref{RE-V}) associated
with the $\Zb_n$ Belavin R-matrix \cite{Fan98}
\begin{eqnarray}
 K^-(u)^s_t=\sum_{i=1}^n \frac{\s(\l_i+\xi-u)}{\s(\l_i+\xi+u)}
 \,\phi^{(s)}_{\l,\l-w\hat{\imath}}(u)
 \,\bar{\phi}^{(t)}_{\l,\l-w\hat{\imath}}(-u). \label{K-matrix}
\end{eqnarray}
The corresponding {\it dual\/} K-matrix $K^+(u)$ which is a
solution to the dual RE (\ref{DRE-V}) has been obtained in
\cite{Yan03}.  With a particular choice of the free boundary
parameters with respect to $K^-(u)$, we introduce the
corresponding dual K-matrix $K^+(u)$
\begin{eqnarray}
 K^+(u)^s_t&=&
 \sum_{i=1}^n \lt\{\prod_{k\neq
 i}\frac{\s((\l_i-\l_k)-w)}{\s(\l_i-\l_k)}\rt\}
 \frac{\s(\l_i+\bar{\xi}+u+
 \frac{nw}{2})}{\s(\l_i+\bar{\xi}-u-\frac{nw}{2})}\no\\[2pt]
 &&\qquad \times \phi^{(s)}_{\l,\l-w\hat{\imath}}(-u)\,
 \tilde{\phi}^{(t)}_{\l,\l-w\hat{\imath}}(u).\label{K-matrix1}
\end{eqnarray}
In (\ref{K-matrix}) and (\ref{K-matrix1}),
$\phi,\,\bar{\phi},\,\tilde{\phi}$ are intertwiners which will be
specified  in section 4. We consider the generic $\{\l_i\}$ such
that $\l_i\neq \l_j\,\, (modulo ~\Zb+\tau\Zb)$ for $i\neq j$. This
condition is necessary for the non-singularity of $K^{\pm}(u)$. It
is convenient to introduce a vector $\l=\sum_{i=1}^n\l_i\e_i$
associated with the boundary parameters $\{\l_i\}$, where
$\{\e_i,~i=1,\cdots,n\}$ is the orthonormal basis of the vector
space $\Cb^n$ such that $\langle \e_i,\e_j\rangle=\d_{ij}$.
Moreover, in the following we always assume that the suffix index
of the parameter $\l_i$ takes value in $\Zb_n$ cyclic group,
namely,
\begin{eqnarray}
 \l_{i\pm n}=\l_i,\qquad i=1,\ldots,n.\label{cylic}
\end{eqnarray}


\section{$A^{(1)}_{n-1}$ SOS R-matrix and face-vertex
correspondence} \label{FV} \setcounter{equation}{0}

The $A_{n-1}$ simple roots $\{\a_i\}$ can be expressed in terms of
the orthonormal basis $\{\e_i\}$  as:
\begin{eqnarray}
 \a_{i}=\e_i-\e_{i+1},\quad i=1,\cdots,n-1,\no
\end{eqnarray}
and the fundamental
weights $\lt\{\L_i~|~i=1,\cdots,n-1\rt\}$ satisfying
$\langle\L_i,~\a_j\rangle=\d_{ij}$ are given by
\begin{eqnarray}
 \L_i=\sum_{k=1}^{i}\e_k-\frac{i}{n}\sum_{k=1}^{n}\e_k. \no
\end{eqnarray}
Set
\begin{eqnarray}
 \hat{\imath}=\e_i-\overline{\e},\quad \overline{\e}=
 \frac{1}{n}\sum_{k=1}^{n}\e_k,\quad i=1,\cdots,n,\quad {\rm
 then}\quad \sum_{i=1}^n\hat{\imath}=0. \label{Vectors}
\end{eqnarray}
For each
dominant weight $\L=\sum_{i=1}^{n-1}a_i\L_{i}$,\ $a_{i}\in
\Zb^+$ (the set of non-negative integers), there exists an
irreducible highest weight finite-dimensional representation
$V_{\L}$ of $A_{n-1}$ with the highest vector $ |\L\rangle$. For
example the fundamental vector representation is $V_{\L_1}$.

Let $\h$ be the Cartan subalgebra of $A_{n-1}$ and $\h^{*}$ be its
dual. A finite dimensional diagonalizable  $\h$-module is a
complex finite dimensional vector space $W$ with a weight
decomposition $W=\oplus_{\mu\in \h^*}W[\mu]$, so that $\h$ acts on
$W[\mu]$ by $x\,v=\mu(x)\,v$, for any $x\in \h,\,v\in\,W[\mu]$.
For example, the fundamental vector representation
$V_{\L_1}=\Cb^n$, the non-zero weight spaces $W[\hat{\imath}]=\Cb
\e_i,~i=1,\cdots,n$.

For a generic $m\in \Cb^n$, define
\begin{eqnarray}
 m_i=\langle m,\e_i\rangle,
 \quad m_{ij}=m_i-m_j=\langle m,\e_i-\e_j\rangle,\quad i,j=1,\cdots,n.
 \label{Def1}
\end{eqnarray}
Let $R(u,m)\in End(\Cb^n\otimes\Cb^n)$ be the R-matrix of the
$A^{(1)}_{n-1}$ SOS model \cite{Jim87},
\begin{eqnarray}
R(u,m)=\sum_{i=1}^{n}
 R^{ii}_{ii}(u,m)E_{ii}
\otimes
 E_{ii} +
 \sum_{i\ne j}
\lt\{R^{ij}_{ij}(u,m)E_{ii}
\otimes
E_{jj}+
 R^{ji}_{ij}(u,m)E_{ji}\otimes
 E_{ij}\rt\}, \label{R-matrix}
\end{eqnarray}
where $E_{ij}$ is the matrix
with elements $(E_{ij})^l_k=\d_{jk}\d_{il}$. The coefficient
functions are
\begin{eqnarray}
 &&R^{ii}_{ii}(u,m)=1,\qquad
 R^{ij}_{ij}(u,m)=\frac{\s(u)\s(m_{ij}-w)}
 {\s(u+w)\s(m_{ij})},\quad i\neq j,\label{Elements1}\\
 && R^{j\,i}_{ij}(u,m)=\frac{\s(w)\s(u+m_{ij})}
 {\s(u+w)\s(m_{ij})},\quad i\neq j,\label{Elements2}
\end{eqnarray}
and  $m_{ij}$
are defined in (\ref{Def1}). The R-matrix satisfies the dynamical
(modified) QYBE
\begin{eqnarray}
&&R_{12}(u_1-u_2,m-wh^{(3)})R_{13}(u_1-u_3,m)
R_{23}(u_2-u_3,m-wh^{(1)})\no\\
&&\qquad =R_{23}(u_2-u_3,m)R_{13}(u_1-u_3,m-wh^{(2)})R_{12}(u_1-u_2,m),
\label{MYBE}
\end{eqnarray}
and the quasi-classical property \begin{eqnarray}
R(u,m)|_{w\rightarrow 0}={\rm id}.\label{quasi1}\end{eqnarray} We
adopt the notation: $R_{12}(u,m-wh^{(3)})$ acts on a tensor
$v_1\otimes v_2 \otimes v_3$ as $R(u,m-w\mu)\otimes id$ if $v_3\in
W[\mu]$. Moreover, the R-matrix satisfies the unitarity and the
modified crossing-unitarity relation.

Let us introduce $n$ intertwiner vectors which are $n$-component
column vectors $\phi_{m,m-w\hat{\jmath}}(u)$ labelled by
$\hat{\jmath}$ ($j=1,\ldots,n$). The  $k$-th element of
$\phi_{m,m-w\hat{\jmath}}(u)$ is given by
\begin{eqnarray}
\phi^{(k)}_{m,m-w\hat{\jmath}}(u)=\theta^{(k)}(u+nm_j).
\label{Intvect}\end{eqnarray}
We remark that  the $n$ intertwiner vectors
$\phi_{m,m-w\hat{\jmath}}(u)$ are linearly independent for a
generic $m\in\Cb^n$.

Using the intertwining vector, one derives the following
face-vertex correspondence relation \cite{Jim87}
\begin{eqnarray}
 &&
 R^B_{12}(u_1-u_2)\, \phi_{m,m-w\hat{\imath}}(u_1)\otimes
 \phi_{m-w\hat{\imath},m-w(\hat{\imath}+\hat{\jmath})}(u_2)\no\\
 &&\qquad = \sum_{k,l}R(u_1-u_2,m)^{kl}_{ij}\,
 \phi_{m-w\hat{l},m-w(\hat{l}+\hat{k})}(u_1)\otimes
 \phi_{m,m-w\hat{l}}(u_2). \label{Face-vertex}
\end{eqnarray}
Then the QYBE (\ref{QYB}) of the $\Zb_n$ Belavin's R-matrix
$R^B(u)$ is equivalent to the dynamical Yang-Baxter equation
(\ref{MYBE}) of the $A^{(1)}_{n-1}$ SOS R-matrix $R(u,m)$. For a
generic $m$, we may introduce other types of intertwiners
$\bar{\phi},~\tilde{\phi}$ satisfying the conditions,
\begin{eqnarray}
 &&\sum_{k=1}^n\bar{\phi}^{(k)}_{m,m-w\hat{\mu}}(u)
 ~\phi^{(k)}_{m,m-w\hat{\nu}}(u)=\d_{\mu\nu},\label{Int1}\\
 &&\sum_{k=1}^n\tilde{\phi}^{(k)}_{m+w\hat{\mu},m}(u)
 ~\phi^{(k)}_{m+w\hat{\nu},m}(u)=\d_{\mu\nu},\label{Int2}
\end{eqnarray}
{}from which one  can derive the relations, \begin{eqnarray}
&&\sum_{\mu=1}^n\bar{\phi}^{(i)}_{m,m-w\hat{\mu}}(u)
~\phi^{(j)}_{m,m-w\hat{\mu}}(u)=\d_{ij},\label{Int3}\\
&&\sum_{\mu=1}^n\tilde{\phi}^{(i)}_{m+w\hat{\mu},m}(u)
~\phi^{(j)}_{m+w\hat{\mu},m}(u)=\d_{ij}.\label{Int4}\end{eqnarray} With the
help of (\ref{Int1})-(\ref{Int4}), we obtain the following
relations from the face-vertex correspondence relation
(\ref{Face-vertex}):
\begin{eqnarray}
 &&\left(\tilde{\phi}_{m+w\hat{k},m}(u_1)\otimes
 {\rm id}\right)R^B_{12}(u_1-u_2) \left(\vTm{\rm
id}\otimes\phi_{m+w\hat{\jmath},m}(u_2)\right)\no\\
 &&\qquad\quad= \sum_{i,l}R(u_1-u_2,m)^{kl}_{ij}\,
 \tilde{\phi}_{m+w(\hat{\imath}+\hat{\jmath}),m+w\hat{\jmath}}(u_1)\otimes
 \phi_{m+w(\hat{k}+\hat{l}),m+w\hat{k}}(u_2),\label{Face-vertex1}\\
 &&\left(\tilde{\phi}_{m+w\hat{k},m}(u_1)\otimes
 \tilde{\phi}_{m+w(\hat{k}+\hat{l}),m+w\hat{k}}(u_2)\right)R^B_{12}(u_1-u_2)\no\\
 &&\qquad\quad= \sum_{i,j}R(u_1-u_2,m)^{kl}_{ij}\,
 \tilde{\phi}_{m+w(\hat{\imath}+\hat{\jmath}),m+w\hat{\jmath}}(u_1)\otimes
 \tilde{\phi}_{m+w\hat{\jmath},m}(u_2),\label{Face-vertex2}\\
 &&\left({\rm id}\otimes
 \bar{\phi}_{m,m-w\hat{l}}(u_2)\right)R^B_{12}(u_1-u_2)
 \left(\vTm\phi_{m,m-w\hat{\imath}}(u_1)\otimes {\rm id}\right)\no\\
 &&\qquad\quad= \sum_{k,j}R(u_1-u_2,m)^{kl}_{ij}\,
 \phi_{m-w\hat{l},m-w(\hat{k}+\hat{l})}(u_1)\otimes
 \bar{\phi}_{m-w\hat{\imath},m-w(\hat{\imath}+\hat{\jmath})}(u_2),\label{Face-vertex3}\\
 &&\left(\bar{\phi}_{m-w\hat{l},m-w(\hat{k}+\hat{l})}(u_1)\otimes
 \bar{\phi}_{m,m-w\hat{l}}(u_2)\right)R^B_{12}(u_1-u_2)\no\\
 &&\qquad\quad= \sum_{i,j}R(u_1-u_2,m)^{kl}_{ij}\,
 \bar{\phi}_{m,m-w\hat{\imath}}(u_1)\otimes
 \bar{\phi}_{m-w\hat{\imath},m-w(\hat{\imath}
 +\hat{\jmath})}(u_2).\label{Face-vertex4}
\end{eqnarray}
The face-vertex correspondence relations (\ref{Face-vertex}) and
(\ref{Face-vertex1})-(\ref{Face-vertex4}) will play an important
role in {\it translating\/}  formulas in the ``vertex form" into
those in the ``face form".

Corresponding to the vertex type K-matrices (\ref{K-matrix}) and
(\ref{K-matrix1}), one introduces the following face type
K-matrices $\K$ and $\tilde{\K}$ \cite{Yan03}
\begin{eqnarray}
 &&\K(\l|u)^j_i=\sum_{s,t}\tilde{\phi}^{(s)}
 _{\l-w(\hat{\imath}-\hat{\jmath}),~\l-w\hat{\imath}}(u)\,K(u)^s_t\,\phi^{(t)}
 _{\l,~\l-w\hat{\imath}}(-u),\label{K-F-1}\\
 &&\tilde{\K}(\l|u)^j_i=\sum_{s,t}\bar{\phi}^{(s)}
 _{\l,~\l-w\hat{\jmath}}(-u)\,\tilde{K}(u)^s_t\,\phi^{(t)}
 _{\l-w(\hat{\jmath}-\hat{\imath}),~\l-w\hat{\jmath}}(u).\label{K-F-2}
\end{eqnarray}
Through straightforward calculations, we find the face type
K-matrices {\it simultaneously\/} have  {\it diagonal\/}
forms\footnote{As will be seen below (see (\ref{parameters})), the
spectral parameter $u$ and the boundary parameter $\xi$ of the
reduced double-row monodromy matrices constructed from $\K(\l|u)$
will be shifted in each step of the nested Bethe Ansatz procedure
\cite{Yan04}. Therefore, it is convenient to specify the
dependence on the boundary parameter $\xi$ of $\K(\l|u)$ in
addition to the spectral parameter $u$. }
\begin{eqnarray}
\K(\l|u)^j_i=\d_i^j\,k(\l|u;\xi)_i,\quad
\tilde{\K}(\l|u)^j_i=\d_i^j\,\tilde{k}(\l|u)_i,\label{Diag-F}
\end{eqnarray}
where functions $k(\l|u;\xi)_i,\,\tilde{k}(\l|u)_i $ are given by
\begin{eqnarray}
 k(\l|u;\xi)_i
 &=&\frac{\s(\l_i+\xi-u)}{\s(\l_i+\xi+u)},\label{k-def}\\
 \tilde{k}(\l|u)_i&=&\lt\{\prod_{k\neq
 i,k=1}^n\frac{\s(\l_{ik}-w)}{\s(\l_{ik})}\rt\}
 \frac{\s(\l_i+\bar{\xi}+u+
 \frac{nw}{2})}{\s(\l_i+\bar{\xi}-u-\frac{nw}{2})}.
 \label{k-def1}
\end{eqnarray}
Moreover, one can check that the matrices $\K(\l|u)$ and
$\tilde{\K}(\l|u)$ satisfy the SOS type reflection equation and
its dual, respectively \cite{Yan03}. Although the K-matrices
$K^{\pm}(u)$ given by (\ref{K-matrix}) and (\ref{K-matrix1}) are
generally non-diagonal (in the vertex form), after the face-vertex
transformations (\ref{K-F-1}) and (\ref{K-F-2}), the face type
counterparts $\K(\l|u)$ and $\tilde{\K}(\l|u)$ become  {\it
simultaneously\/} diagonal. This fact enables the authors  to
apply the generalized algebraic Bethe Ansatz method developed in
\cite{Yan04} for SOS type integrable models to diagonalize the
transfer matrix $\t(u)$ (\ref{trans}).


\section{Algebraic Bethe Ansatz} \label{BAE} \setcounter{equation}{0}

By means of (\ref{Int3}), (\ref{Int4}), (\ref{K-F-2}) and
(\ref{Diag-F}), the transfer matrix $\t(u)$ (\ref{trans}) can be
recast  into the following face type form:
\begin{eqnarray}
 \hspace{-1.2truecm}\t(u)\hspace{-0.22truecm}&=&
 \hspace{-0.22truecm}tr(K^+(u)\mathbb{T}(u))\no\\
 \hspace{-0.22truecm}&=&\hspace{-0.22truecm}
 \sum_{\mu,\nu}tr\hspace{-0.12truecm}\lt(\!K^+(u)
 \phi_{\l-w(\hat{\mu}-\hat{\nu}),
 \l-w\hat{\mu}}(u)\tilde{\phi}_{\l-w(\hat{\mu}-\hat{\nu}),
 \l-w\hat{\mu}}(u)\mathbb{T}(u)
 \phi_{\l, \l-w\hat{\mu}}(\hspace{-0.08truecm}-\hspace{-0.08truecm}u)
 \bar{\phi}_{\l, \l-w\hat{\mu}}(\hspace{-0.08truecm}-\hspace{-0.08truecm}u)
 \hspace{-0.12truecm}\rt)\no\\
 \hspace{-0.22truecm}&=&\hspace{-0.22truecm}
 \sum_{\mu,\nu}\bar{\phi}_{\l, \l-w\hat{\mu}}(-u)K^+(u)
 \,\phi_{\l-w(\hat{\mu}-\hat{\nu}),\l-w\hat{\mu}}(u)~
 \tilde{\phi}_{\l-w(\hat{\mu}-\hat{\nu}),
 \l-w\hat{\mu}}(u)~\mathbb{T}(u)\phi_{\l, \l-w\hat{\mu}}(-u)\no\\
 \hspace{-0.22truecm}&=&\hspace{-0.22truecm}
 \sum_{\mu,\nu}\tilde{\K}(\l|u)_{\nu}^{\mu}\,\T(\l|u)^{\nu}_{\mu}=
 \sum_{\mu}\tilde{k}(\l|u)_{\mu}\,\T(\l|u)^{\mu}_{\mu}.
 \label{De1}
\end{eqnarray}
Here we have introduced the face-type double-row
monodromy matrix $\T(\l|u)$,
\begin{eqnarray}
 \T(\l|u)^{\nu}_{\mu}&=&\tilde{\phi}_{\l-w(\hat{\mu}-\hat{\nu}),
 \l-w\hat{\mu}}(u)~\mathbb{T}(u)\,\phi_{\l,
 \l-w\hat{\mu}}(-u)\no\\
 &\equiv&
 \sum_{i,j}\tilde{\phi}^{(j)}_{\l-w(\hat{\mu}-\hat{\nu}),
 \l-w\hat{\mu}}(u)~\mathbb{T}(u)^j_i\,\phi^{(i)}_{\l,
 \l-w\hat{\mu}}(-u).\label{Mon-F}
\end{eqnarray}
This face-type double-row monodromy matrix can  be expressed in
terms of the face type R-matrix $R(u,\l)$ (\ref{R-matrix}) and the
K-matrix $\K(\l|u)$ (\ref{K-F-1}) \cite{Yan04}. Moreover from
(\ref{Relation-Re}), (\ref{Face-vertex}) and (\ref{Int4}) one may
derive the following exchange relations among
$\T(m|u)^{\nu}_{\mu}$:
\begin{eqnarray}
 &&\sum_{i_1,i_2}\sum_{j_1,j_2}~
 R(u_1-u_2,m)^{i_0\,j_0}_{i_1\,j_1}\,\T(m+w(\hat{\jmath}_1+\hat{\imath}_2)|u_1)
 ^{i_1}_{i_2}\no\\
 &&~~~~~~~~\times R(u_1+u_2,m)^{j_1\,i_2}_{j_2\,i_3}\,
 \T(m+w(\hat{\jmath}_3+\hat{\imath}_3)|u_2)^{j_2}_{j_3}\no\\[2pt]
 &&~~=\sum_{i_1,i_2}\sum_{j_1,j_2}~
 \T(m+w(\hat{\jmath}_1+\hat{\imath}_0)|u_2)
 ^{j_0}_{j_1}\,R(u_1+u_2,m)^{i_0\,j_1}_{i_1\,j_2}\no\\
 &&~~~~~~~~\times\T(m+w(\hat{\jmath}_2+\hat{\imath}_2)|u_1)^{i_1}_{i_2}\,
 R(u_1-u_2,m)^{j_2\,i_2}_{j_3\,i_3}.\label{RE-F}
\end{eqnarray}

Following \cite{Yan04} and motivated by our recent work
\cite{Yan07} for the open XXZ chain, we introduce $n$ sets of
operators $\{\A^{(s)},\B^{(s)},\C^{(s)},\D^{(s)}\}$, labelled by
$s=1,\ldots,n$, as follows:
\begin{eqnarray}
\hspace{-0.28truecm}\A^{(s)}(m|u) \hspace{-0.28truecm}&=&
\hspace{-0.28truecm}\T(m|u)^s_s,\quad \B^{(s)}_i(m|u)=
 \frac{\T(m|u)^s_i}{\s(m_{is})},
 \quad \C^{(s)}_i(m|u)=
 \frac{\T(m|u)^i_s}{\s(m_{si})},\quad i\neq s,\label{Def-AB}\\
\hspace{-0.28truecm}
{\D^{(s)}}^j_i(m|u)\hspace{-0.28truecm}&=&\hspace{-0.28truecm}
\frac{\s(m_{js}-\d_{ij}w)}{\s(m_{is})}
 \left\{\T(m|u)^j_i-\d^j_iR(2u,m+w\hat{s})^{j\,\,s}_{s\,j}\,\A^{(s)}
 (m|u)\right\}
 ,\quad i,j\neq s. \label{Def-D}
\end{eqnarray} Some remarks are in order. Among the $n$ sets of
operators, the first set (corresponding to $s=1$) is the very one
which was used in \cite{Yan04,Yan04-4} to construct the algebraic
Bethe Ansatz. Such algebraic Bethe Ansatz based on the first set
of operators only gives rise to the first set of eigenvalues and
associated Bethe Ansatz equations. In order to find the complete
sets of eigenvalues, we find that the whole $n$ sets of operators
$\{\A^{(s)},\B^{(s)},\C^{(s)},\D^{(s)}|\,s=1,\ldots, n\}$ are
needed.

After tedious calculations analogous to those in \cite{Yan04}, we
have found the commutation relations among $\A^{(s)}(m|u)$,
$\D^{(s)}(m|u)$ and $\B^{(s)}(m|u)$ from (\ref{RE-F}). Here we
give those which are relevant for our purpose
\begin{eqnarray}
&&\hspace{-0.28truecm}\A^{(s)}(m|u)\B^{(s)}_i(m\hspace{-0.1truecm}
+\hspace{-0.1truecm}w(\hat{\imath}-\hat{s})|v)\no\\
&&\quad=\hspace{-0.1truecm}\frac{\s(u+v)\s(u-v-w)}{\s(u+v+w)\s(u-v)}
\,\B^{(s)}_i(m\hspace{-0.1truecm}+\hspace{-0.1truecm}
w(\hat{\imath}-\hat{s})|v)\A^{(s)}(m\hspace{-0.1truecm}+\hspace{-0.1truecm}
w(\hat{\imath}-\hat{s})|u)\no\\
&&\quad\quad-
\frac{\s(w)\s(2v)}{\s(u\hspace{-0.1truecm}-\hspace{-0.1truecm}
v)\s(2v\hspace{-0.1truecm}+\hspace{-0.1truecm}w)}
\frac{\s(u\hspace{-0.1truecm}-\hspace{-0.1truecm}v\hspace{-0.1truecm}
-\hspace{-0.1truecm}m_{si}\hspace{-0.1truecm}+\hspace{-0.1truecm}w)}
{\s(m_{si}-w)}
\,\B^{(s)}_i(m\hspace{-0.1truecm}+\hspace{-0.1truecm}
w(\hat{\imath}\hspace{-0.1truecm}-\hspace{-0.1truecm}\hat{s})|u)
\A^{(s)}(m\hspace{-0.1truecm}+\hspace{-0.1truecm}
w(\hat{\imath}\hspace{-0.1truecm}-\hspace{-0.1truecm}\hat{s})|v)\no\\
&&\quad\quad-
\frac{\s(w)}{\s(u\hspace{-0.1truecm}+\hspace{-0.1truecm}
v\hspace{-0.1truecm}+\hspace{-0.1truecm}w)}\sum_{\a\neq s}
\frac{\s(u\hspace{-0.1truecm}+\hspace{-0.1truecm}v
\hspace{-0.1truecm}+\hspace{-0.1truecm}m_{\a
s}\hspace{-0.1truecm}+\hspace{-0.1truecm}2w)}{\s(m_{\a s}+w)}
\,\B^{(s)}_{\a}(m\hspace{-0.1truecm}+\hspace{-0.1truecm}
w(\hat{\a}\hspace{-0.1truecm}-\hspace{-0.1truecm}\hat{s})|u)
{\D^{(s)}}^{\a}_i(m\hspace{-0.1truecm}+\hspace{-0.1truecm}
w(\hat{\imath}\hspace{-0.1truecm}-\hspace{-0.1truecm}\hat{s})|v),\no\\
&&\qquad\qquad\qquad\qquad\qquad\qquad\qquad\qquad  i\neq s,\label{Rel-1}\\[6pt]
&&{\D^{(s)}}^k_i(m|u)\B^{(s)}_j(m+w(\hat{\jmath}-\hat{s})|v)\no\\
&&\qquad=
\frac{\s(u-v+w)\s(u+v+2w)}{\s(u-v)\s(u+v+w)}\no\\
&&\qquad\qquad\qquad\times\lt\{\sum_{\a_1,\a_2,\b_1,\b_2\neq s}
R(u+v+w,m-w\hat{\imath})^{k\,\,\,\,\b_2}_{\a_2\,\b_1}
R(u-v,m+w\hat{\jmath})^{\b_1\,\a_1}_{j\,\,\,\,i}\rt.\no\\
&&\qquad\qquad\qquad\qquad
\times\lt.\B^{(s)}_{\b_2}(m+w(\hat{k}+\hat{\b}_2-\hat{\imath}-\hat{s})|v)
{\D^{(s)}}^{\a_2}_{\a_1}(m+w(\hat{\jmath}-\hat{s})|u)\rt\}\no\\
&&\qquad\quad-\frac{\s(w)\s(2u+2w)}{\s(u-v)\s(2u+w)}\lt\{
\sum_{\a,\b\neq s}\frac{\s(u-v+m_{s\a}-w)}{\s(m_{s\a}-w)}
\,R(2u+w,m-w\hat{\imath})^{k\,\b}_{\a\,i}\rt.
\no\\
&&\qquad\qquad\qquad\qquad\times\lt.
\B^{(s)}_{\b}(m+w(\hat{k}+\hat{\b}-\hat{\imath}-\hat{s})|u)
{\D^{(s)}}^{\a}_{j}(m+w(\hat{\jmath}-\hat{s})|v)\rt\}\no\\
&&\qquad\quad+\frac{\s(w)\s(2v)\s(2u+2w)}{\s(u+v+w)\s(2v+w)\s(2u+w)}\no\\[2pt]
&&\qquad\qquad\qquad\times\lt\{ \sum_{\a\neq
s}\frac{\s(u+v+m_{sj})}{\s(m_{sj}-w)}
\,R(2u+w,m-w\hat{\imath})^{k\,\a}_{j\,i}\rt.
\no\\
&&\qquad\qquad\qquad\qquad\times\lt.
\B^{(s)}_{\a}(m+w(\hat{k}+\hat{\a}-\hat{\imath}-\hat{s})|u)
\A^{(s)}(m+w(\hat{j}-\hat{s})|v)\rt\},\no\\
&&\qquad\qquad\qquad\qquad\qquad\qquad\qquad\qquad i,j,k\neq s,\label{Rel-2}\\[6pt]
&&\B^{(s)}_i(m+w(\hat{\imath}-\hat{s})|u)
\B^{(s)}_j(m+w(\hat{\imath}+\hat{\jmath}-2\hat{s})|v)\no\\
&&\qquad=\hspace{-0.1truecm}\sum_{\a,\b\neq
s}\hspace{-0.1truecm}R(u\hspace{-0.08truecm}-\hspace{-0.08truecm}v,
m\hspace{-0.08truecm}-\hspace{-0.08truecm}2w\hat{s})^{\b\,\a}_{j\,i}
\B^{(s)}_{\b}(m\hspace{-0.08truecm}+\hspace{-0.08truecm}
w(\hat{\b}\hspace{-0.08truecm}-\hspace{-0.08truecm}\hat{s})|v)
\B^{(s)}_{\a}(m\hspace{-0.08truecm}+\hspace{-0.08truecm}
w(\hat{\a}\hspace{-0.08truecm}+\hspace{-0.08truecm}\hat{\b}\hspace{-0.08truecm}
-\hspace{-0.08truecm}2\hat{s})|u),\no\\
&&\qquad\qquad\qquad\qquad\qquad\qquad\qquad\qquad i,j\neq
s.\label{Rel-3}
\end{eqnarray}
For the special case of $s=1$,  the above commutation relations
(\ref{Rel-1})-(\ref{Rel-3}) recover those in \cite{Yan04,Yan04-4}.

In order to apply the algebraic Bethe Ansatz method, in addition
to the relevant commutation relations (\ref{Rel-1})-(\ref{Rel-3}),
one needs to construct a reference state associated with each $s$,
which is the common eigenstate of the operators $\A^{(s)}$,
${\D^{(s)}}^i_i$ and is annihilated by the operators $\C^{(s)}_i$.
In contrast to the trigonometric and rational cases with {\it
diagonal\/} $K^{\pm}(u)$ \cite{Skl88}, the usual highest-weight
state
\begin{eqnarray}
 \lt(\begin{array}{l}1\\0\\\vdots\end{array}\rt)\otimes\cdots\otimes
 \lt(\begin{array}{l}1\\0\\\vdots\end{array}\rt),\no
\end{eqnarray}
is no longer the pseudo-vacuum state. However, after the
face-vertex transformations (\ref{K-F-1}) and (\ref{K-F-2}), the
face type K-matrices $\K(\l|u)$ and $\tilde{\K}(\l|u)$ {\it
simultaneously\/} become diagonal. This suggests that one can
translate the $\Zb_n$ Belavin model  with  non-diagonal K-matrices
into the corresponding SOS model with {\it diagonal} K-matrices
$\K(\l|u)$ and $\tilde{\K}(\l|u)$ given by
(\ref{K-F-1})-(\ref{K-F-2}). Then one can construct the
pseudo-vacuum in the ``face language" and use the algebraic Bethe
Ansatz method  to diagonalize the transfer matrix.

One of the reference states corresponding to the case of $s=1$ (or
the first one of the $n$ reference states (\ref{Vac}) below) was
found in \cite{Yan04} and yields the first set of eigenvalues of
the transfer matrix. Here, we give the complete $n$ reference
states $\{|\O^{(s)}(\l)\rangle|s=1,\ldots,n\}$. For each $s$, we
propose\footnote{Such states played an important role in
constructing extra centers of the elliptic algebra at roots of
unity \cite{Yan052}.}
\begin{eqnarray}
 |\O^{(s)}(\l)\rangle=\phi_{\l-(N-1)w\hat{s},\l-Nw\hat{s}}(-z_1)\otimes
 \phi_{\l-(N-2)w\hat{s},\l-(N-1)w\hat{s}}(-z_{2})\cdots\otimes
 \phi_{\l,\l-w\hat{s}}(-z_N).\label{Vac}
\end{eqnarray}
These states ($s=1,\ldots,n$) depend on the boundary parameters
$\{\l_i\}$ and the inhomogeneous parameters $\{z_j\}$, but not on
the boundary parameters $\xi$ and $\bar{\xi}$. We find that the
states given by (\ref{Vac}) are {\it exactly} the reference states
in the following sense,
\begin{eqnarray}
\A^{(s)}(\l-Nw\hat{s}|u)|\O^{(s)}(\l)\rangle&=&k(\l|u;\xi)_s
|\O^{(s)}(\l)\rangle,\label{A}\\
{\D^{(s)}}^i_j(\l-Nw\hat{s}|u)|\O^{(s)}(\l)\rangle&=&\d^i_j
f^{(s)}(u)\,k(\l|u+\frac{w}{2};\xi-\frac{w}{2})_j\no\\
&&\hspace{-1mm}\times\!\lt\{\prod_{k=1}^N
\frac{\s(u+z_k)\s(u-z_k)}{\s(u+z_k+w)\s(u-z_k+w)}\!\rt\}\!
|\O^{(s)}(\l)\rangle,\no\\
&&\qquad\quad i,j\neq s\label{D}\\
\C^{(s)}_i(\l-Nw\hat{s}|u)|\O^{(s)}(\l)\rangle&=&0,\qquad i\neq s\\
\B^{(s)}_i(\l-Nw\hat{s}|u)|\O^{(s)}(\l)\rangle&\neq& 0,\qquad
i\neq s.
\end{eqnarray} Here $f^{(s)}(u)$ is given by  \begin{eqnarray}
f^{(s)}(u)= \frac{\s(2u)\s(\l_s+u+w+\xi)}{\s(2u+w)\s(\l_s+u+\xi)}.
\end{eqnarray}

In order to apply the algebraic Bethe Ansatz method to diagonalize
the transfer matrix, we need to assume   $N=n\times l$ with $l$
being a positive integer \cite{Yan04}. For convenience, let us
introduce a set of integers: \begin{eqnarray} N_i=(n-i)\times
l,~~i=0,1,\cdots,n-1,\label{Integer} \end{eqnarray} and
$\frac{n(n-1)}{2}l$ complex parameters
$\{v^{(i)}_k|~k=1,2,\cdots,N_{i+1},~i=0,1,\cdots,n-2\}$. As in the
usual nested Bethe Ansatz method, the parameters $\{v^{(i)}_k\}$
will be used to specify the eigenvectors of the corresponding
reduced transfer matrices.  They will be constrained later by
Bethe Ansatz equations. For convenience, we adopt the following
convention:
\begin{eqnarray}
 v_k=v^{(0)}_k,~k=1,2,\cdots, N_1.
\end{eqnarray}
We will seek the common eigenvectors (i.e. the so-called Bethe
states) of the transfer matrix by acting the creation operators
$\B_i^{(s)}$ on the reference state $|\O^{(s)}(\l)\rangle$
\begin{eqnarray}
\hspace{-0.8truecm}|v_1,\cdots,v_{N_1}\rangle^{(s)}
\hspace{-0.18truecm}&=&\hspace{-0.28truecm}\sum_{i_1,\cdots,i_{N_1}\neq
s} F^{i_1,i_2,\cdots,i_{N_1}}
\B^{(s)}_{i_1}(\l\hspace{-0.08truecm}+\hspace{-0.08truecm}
w(\hat{\imath}_1\hspace{-0.08truecm}-\hspace{-0.08truecm}\hat{s})|v_{1})
\B^{(s)}_{i_2}(\l\hspace{-0.08truecm}+\hspace{-0.08truecm}
w(\hat{\imath}_1\hspace{-0.08truecm}+\hspace{-0.08truecm}\hat{\imath}_2
\hspace{-0.08truecm}-\hspace{-0.08truecm}2\hat{s})|v_2)\no\\
&&\qquad\qquad\qquad\times\cdots\B^{(s)}_{i_{N_1}}
(\l+w\sum_{k=1}^{N_1}\hat{\imath}_k-wN_1\hat{s}|v_{N_1})
|\O^{(s)}(\l)\rangle.\label{Eigenstate}
\end{eqnarray}
The indices in the above
equation should satisfy the following condition: the number of
$i_k=j$, denoted by $\#(j)$, is $l$, i.e.
\begin{eqnarray}
  \#(j)=l, ~~~j\neq s.\label{Restriction}
\end{eqnarray} Then (\ref{Vectors}) and the above restriction
(\ref{Restriction}) imply
\begin{eqnarray}
 \l+w\sum_{k=1}^{N_1}\hat{\imath}_k-wN_1\hat{s}=\l+wl\sum_{j\neq
 s}^n\hat{\jmath}-wN_1\hat{s}=\l-w(l+N_1)\hat{s}=\l-wN\hat{s},
\end{eqnarray} which is crucial for the diagonalization of the
transfer matrix in the remaining part of the paper.

With the help of (\ref{De1}), (\ref{Def-AB}) and (\ref{Def-D}) we
rewrite the transfer matrix (\ref{trans}) in terms of the
operators $\A^{(s)}$ and ${\D^{(s)}}^i_i$
\begin{eqnarray}
 \t(u)&=&\sum_{\nu=1}^n\tilde{k}(\l|u)_{\nu}\,\T(\l|u)^{\nu}_{\nu}
 \no\\
 &=&\tilde{k}(\l|u)_s\,\A^{(s)}(\l|u)+\sum_{i\neq s}\tilde{k}(\l|u)_i
 \,\T(\l|u)^i_i \no\\
 &=&\tilde{k}(\l|u)_s\,\A^{(s)}(\l|u)+ \sum_{i\neq s}\tilde{k}(\l|u)_i
 \,R(2u,\l+w\hat{s})^{is}_{si}\,\A^{(s)}(\l|u)\no\\
 &&
 \qquad+ \sum_{i\neq s}
 \tilde{k}(\l|u)_i
 \left(\vTm\T(\l|u)^i_i- R(2u,\l+w\hat{s})^{is}_{si}
 \A^{(s)}(\l|u)\right)\no\\
 &=& \sum_{i=1}^n\tilde{k}(\l|u)_i
 R(2u,\l+w\hat{s})^{is}_{si}\,\A^{(s)}(\l|u)\no\\
 &&\qquad+ \sum_{i\neq s}\tilde{k}^{(1)}(\l|u\hspace{-0.08truecm}+
 \hspace{-0.08truecm}\frac{w}{2})_i
 \frac{\s(\l_{is}\hspace{-0.08truecm}-\hspace{-0.08truecm}w)}
 {\s(\l_{is})}\hspace{-0.08truecm}
 \left(\vTm \hspace{-0.08truecm}
 \T(\l|u)^i_i\hspace{-0.08truecm}-\hspace{-0.08truecm}
 R(2u,\l\hspace{-0.08truecm}+\hspace{-0.08truecm}w\hat{s})^{is}_{si}
 \A^{(s)}(\l|u)\hspace{-0.08truecm}\right)\no\\
 &=&\a_s^{(1)}(u)\,\A^{(s)}(\l|u)+\sum_{i\neq s}
 \tilde{k}^{(1)}(\l|u+\frac{w}{2})_i\,{\D^{(s)}}^i_i(\l|u).
 \label{trans1}
\end{eqnarray}
Here we have used (\ref{Def-D}) and introduced the function
$\a_s^{(1)}(u)$,
\begin{eqnarray}
 \a_s^{(1)}(u)=\sum_{i=1}^n\tilde{k}(\l|u)_i\,
 R(2u,\l+w\hat{s})^{is}_{si},\label{function-a}
\end{eqnarray}
and the reduced
 K-matrix $\tilde{\K}^{(1)}(\l|u)$ with elements given by
\begin{eqnarray}
\tilde{\K}^{(1)}(\l|u)^j_i&=&
 \d^j_i\,\tilde{k}^{(1)}(\l|u)_i,\qquad\qquad\qquad
 i,j\neq s\label{Reduced-K1}\\
\tilde{k}^{(1)}(\l|u)_i&=& \lt\{\prod_{k\neq
 i,s}^n\frac{\s(\l_{ik}-w)}{\s(\l_{ik})}\rt\}
 \frac{\s(\l_i+\bar{\xi}+u+
 \frac{(n-1)w}{2})}{\s(\l_i+\bar{\xi}-u-\frac{(n-1)w}{2})},\quad i\neq s.
 \label{Reduced-K2}
\end{eqnarray}

Using the technique developed in \cite{Yan04}, after tedious
calculations, we find that with the coefficients
$F^{i_1,i_2,\cdots,i_{N_1}}$ in (\ref{Eigenstate}) properly
chosen,  the Bethe state $|v_1,\cdots,v_{N_1}\rangle^{(s)} $
becomes the eigenstate of the transfer matrix (\ref{trans}),
\begin{eqnarray}
 \t(u)|v_1,\cdots,v_{N_1}\rangle^{(s)}=\L_s(u;\xi,\{v_k\})
 |v_1,\cdots,v_{N_1}\rangle^{(s)},
\end{eqnarray} provided that the parameters
$\{v^{(i)}_k|~k=1,2,\cdots,N_{i+1},~i=0,1,\cdots,n-2\}$ satisfy
the following Bethe Ansatz equations:
\begin{eqnarray}
 &&\hspace{-10pt}\b_s^{(1)}(v_j)\frac{\s(2v_j+w)}
 {\s(2v_j+2w)} \prod_{k\ne j,k=1}^{N_1}
 \frac{\s(v_j+v_k)\s(v_j-v_k-w)} {\s(v_j+v_k+2w)\s(v_j-v_k+w)}
 \no\\
 &&\qquad\qquad = \prod_{k=1}^{N} \frac{\s(v_j+z_k)\s(v_j-z_k)}
 {\s(v_j+z_k+w)\s(v_j-z_k+w)}
 \,\L_s^{(1)}(v_j+\frac{w}{2};\xi-\frac{w}{2},\{v^{(1)}_a\}),\qquad\qquad
 \label{BA1}\\
 &&\hspace{-10pt}\b_s^{(i+1)}(v^{(i)}_j)\frac{
 \s(2v^{(i)}_j+w)}
 {\s(2v^{(i)}_j+2w)}\!\!
 \prod_{k\ne j,k=1}^{N_{i+1}}
 \frac{\s(v^{(i)}_j+v^{(i)}_k)\s(v^{(i)}_j-v^{(i)}_k-w)}
 {\s(v^{(i)}_j+v^{(i)}_k+2w)\s(v^{(i)}_j-v^{(i)}_k+w)}\no\\
 &&\qquad\qquad = \prod_{k=1}^{N_i}\hspace{-0.08truecm}
 \frac{\s(v^{(i)}_j+z^{(i)}_k)\s(v^{(i)}_j-z^{(i)}_k)}
 {\s(v^{(i)}_j\hspace{-0.08truecm}+\hspace{-0.08truecm}z^{(i)}_k
 \hspace{-0.08truecm}+\hspace{-0.08truecm}w)
 \s(v^{(i)}_j\hspace{-0.08truecm}-\hspace{-0.08truecm}z^{(i)}_k
 \hspace{-0.08truecm}+\hspace{-0.08truecm}w)}
 \L_s^{(i+1)}(v^{(i)}_j\hspace{-0.08truecm}+\hspace{-0.08truecm}
 \frac{w}{2};\xi^{(i)}\hspace{-0.08truecm}-\hspace{-0.08truecm}
 \frac{w}{2},\{v^{(i+1)}_a\})
 ,\no\\
 &&\qquad\qquad\qquad\qquad\qquad\qquad i=1,\cdots,n-2.
 \label{BA2}
\end{eqnarray} Here $\{\b^{(i)}_s(u)|i=1,\dots,n-1\}$ are functions
given in Appendix A,
$\{\L_s^{(i)}(u;\xi,\{v^{(i)}_{k}\})|i=1,\ldots,n-1\}$ are the
eigenvalues (given in Appendix A)  of the reduced transfer
matrices in the nested Bethe Ansatz process, and the reduced
boundary parameters $\{\xi^{(i)}\}$ and inhomogeneous parameters
$\{z^{(i)}_k\}$ are given by
\begin{eqnarray}
 \xi^{(i+1)}=\xi^{(i)}-\frac{w}{2},
 \qquad z^{(i+1)}_k=v^{(i)}_k+\frac{w}{2},
 \qquad i=0,\cdots,n-2.\label{parameters}
\end{eqnarray} In the above we have adopted the
convention: $\xi=\xi^{(0)}$, $z^{(0)}_k=z_k$. The corresponding
eigenvalue  $\L_s(u;\xi,\{v_k\})$ is given by
 \begin{eqnarray}
 \hspace{-0.42truecm}
 \L_s(u;\xi,\{v_k\})&=&\b_s^{(1)}(u)\,
 \frac{\s(\l_s+\xi+u+w)}{\s(\l_s+\xi+u)}\,
 \prod_{k=1}^{N_1}\frac{\s(u+v_k)\s(u-v_k-w)}
 {\s(u+v_k+w)\s(u-v_k)}\no\\
 &&+\frac{\s(2u)\s(\l_s+u+w+\xi)} {\s(2u+w)\s(\l_s+u+\xi)}
 \prod_{k=1}^{N_1}\frac{\s(u-v_k+w)\s(u+v_k+2w)}
 {\s(u-v_k)\s(u+v_k+w)}\no\\
 &&\ \ \times \prod_{k=1}^{N_0}\frac{\s(u+z_k)\s(u-z_k)}
 {\s(u+z_k+w)\s(u-z_k+w)}
 \L_s^{(1)}(u\hspace{-0.08truecm}+\hspace{-0.08truecm}
 \frac{w}{2};\xi\hspace{-0.08truecm}-\hspace{-0.08truecm}
 \frac{w}{2},\{v_i^{(1)}\}).\label{Eigenvalue1}
\end{eqnarray}

It is easy to check that the first set of eigenvalues
$\L_1(u;\xi,\{v^{(i)}_{k}\})$ and the corresponding Bethe Ansatz
equations are exactly those found in \cite{Yan04}. However, the
rest $n-1$ sets of eigenvalues
$\{\L_s(u;\xi,\{v_{k}\})|s=2,\ldots,n-1\}$ and the associated
Bethe Ansatz equations are new ones. For the special case of
$n=2$, which corresponds to the open XYZ spin chain, we find that
the two sets of eigenvalues $\{\L_s(u;\xi,\{v_{k}\})|s=1,2\}$,
after rescaling of an overall factor due to different
normalizations of the R- and K-matrices, recover those in
\cite{Yan062} obtained  by directly solving the $T$-$Q$ relation
(or functional Bethe Ansatz). As shown in \cite{Yan062}, these two
sets of eigenvalues give the complete spectrum of the transfer
matrix of the open XYZ spin chain. As a consequence, the
corresponding two sets of Bethe states
$\{|v_1,\cdots,v_{N_1}\rangle^{(s)}|s=1,2\}$ together constitute
the complete  eigenstates of the transfer matrix of the open XYZ
model. Therefore, it is expected that the $n$ sets of eigenvalues
$\{\L_s(u;\xi,\{v_{k}\})|s=1,\ldots,n \}$ (\ref{Eigenvalue1})
[resp. Bethe states
$\{|v_1,\cdots,v_{N_1}\rangle^{(s)}|s=1,\ldots, n\}$
(\ref{Eigenstate}) and (\ref{BA1})-(\ref{BA2})] together give rise
to the complete spectrum [resp. the complete eigenstates] of the
transfer matrix (\ref{trans}) of the open $\Zb_n$ Belavin model.


\section{Result for the associated Gaudin model}
\label{Gou} \setcounter{equation}{0}

As will be seen from the definitions of the intertwiners
(\ref{Intvect}), (\ref{Int1}) and (\ref{Int2}), specialized to
$m=\l$, $\phi_{\l,\l-w\hat{\imath}}(u)$ and
$\bar{\phi}_{\l,\l-w\hat{\imath}}(u)$ do not depend on $w$ while
$\tilde{\phi}_{\l,\l-w\hat{\imath}}(u)$ does. Consequently, the
K-matrix $K^-(u)$ does not depend on the crossing parameter $w$,
 but $K^+(u)$ does. So we use the
convention:
\begin{eqnarray}
 K(u)=\lim_{w\rightarrow
 0}K^-(u)=K^-(u).\label{Conv1}
\end{eqnarray}
We further assume that the parameter $\bar{\xi}$ has the following
behavior\footnote{In \cite{Yan041,Yan04-4,Yan07}, a special case
of $\bar{\xi}=\xi$ was studied. The generalization to the case
 with nonvanishing $\d$ is straightforward.} as $w\rightarrow 0$,
\begin{eqnarray}
 \bar{\xi}=\xi+w\d+O(w^2),\label{Restriction-1}
\end{eqnarray} with a constant $\d$. It implies that
\begin{eqnarray}
 \lim_{w\rightarrow 0}\,\bar{\xi}=\xi.\no
\end{eqnarray} Then
the K-matrices satisfy the following relation
\begin{eqnarray}
\lim_{w\rightarrow 0}\{K^+(u)\,K^-(u)\}=\lim_{w\rightarrow
0}\{K^+(u)\}K(u)={\rm id}.\label{ID-1}
\end{eqnarray}

Let us introduce  the elliptic  Gaudin operators  $\{H_j
|j=1,2,\cdots,N\}$ associated with the inhomogeneous $\Zb_n$
Belavin  model with open boundaries specified by the generic
K-matrices (\ref{K-matrix}) and (\ref{K-matrix1}):
\begin{eqnarray}
H_j=\G_j(z_j)+\sum_{k\neq
 j}^{N}r_{kj}(z_j-z_k)+K^{-1}_j(z_j)\lt\{\sum_{k\neq
 j}^{N}r_{jk}(z_j+z_k)\rt\}K_j(z_j),\label{Ham}
\end{eqnarray}
where $\G_j(u)=\frac{\partial}{\partial
w}\{\bar{K}_j(u)\}|_{w=0}K_j(u)$, $j=1,\cdots,N,$ with
$\bar{K}_j(u)=tr_0\lt\{K^+_0(u)R^B_{0j}(2u)P_{0j}\rt\}$. Here
$\{z_j\}$ are the inhomogeneous parameters of the inhomogeneous
$\Zb_n$ Belavin model and $r(u)$ is given by (\ref{r-matrix}). For
a generic choice of the boundary parameters
$\{\l_1,\,\cdots,\l_n,\,\bar{\xi}\}$, $\G_j(u)$ is a {\it
non-diagonal\/} matrix.

Following \cite{Hik95,Skl96}, the elliptic  Gaudin operators
(\ref{Ham}) are obtained by expanding the double-row transfer
matrix (\ref{trans}) at the point $u=z_j$ around $w=0$:
\begin{eqnarray}
 \t(z_j)&=&\t(z_j)|_{w=0}+w
 H_j+O(w^2),\quad j=1,\cdots,N, \label{trans-2}\\
 H_j&=&\frac{\partial}{\partial w}\t(z_j)|_{w=0}\,  .\label{Eq-1}
\end{eqnarray}
The relations (\ref{quasi}) and (\ref{ID-1}) imply that the first
term $\t(z_j)|_{w=0}$ in the expansion (\ref{trans-2}) is equal to
an identity, namely,
\begin{eqnarray}
 \t(z_j)|_{w=0}={\rm id}.\label{First}
\end{eqnarray}
Then the commutativity of the transfer matrices $\{\t(z_j)\}$
(\ref{Com-2}) for a generic $w$ implies
\begin{eqnarray}
[H_j,H_k]=0,\quad i,j=1,\cdots,N.\label{Com-1}
\end{eqnarray} Thus the elliptic
Gaudin system defined by (\ref{Ham}) is integrable. Moreover, the
relation (\ref{Eq-1}) between $\{H_j\}$ and $\{\t(z_j)\}$ and the
fact that the first term on the r.h.s. of (\ref{trans-2}) is
identity operator enable us to extract the eigenstates of the
elliptic Gaudin operators and the corresponding eigenvalues from
the results obtained in the previous section.

Using (\ref{parameters}), (\ref{Eigenvalue2}), (\ref{Eigenvalue3})
and (\ref{alpha-functions})-(\ref{lemada-functions}), the Bethe
Ansatz equations (\ref{BA1}) and (\ref{BA2}) become, respectively,
\begin{eqnarray}
 &&\hspace{-24pt} \b_s^{(i+1)}(v^{(i)}_j)\frac{\s(2v^{(i)}_j+w)}{\s(2v^{(i)}_j+2w)}
 \prod_{k\neq j,k=1}^{N_{i+1}}\frac{\s(v^{(i)}_j+v^{(i)}_k)
 \s(v^{(i)}_j-v^{(i)}_k-w)}{\s(v^{(i)}_j+v^{(i)}_k+2w)
 \s(v^{(i)}_j-v^{(i)}_k+w)}\no\\
 && ~~~~=\b_s^{(i+2)}(v^{(i)}_j+\frac{w}{2})\,
 \frac{\s(\l_{i+s+1}+\x^{(i+1)}+v^{(i)}_j+\frac{3}{2}w)}
 {\s(\l_{i+s+1}+\x^{(i+1)}+v^{(i)}_j+\frac{1}{2}w)}\no\\
 &&\qquad\qquad\qquad\qquad\times
 \prod_{k=1}^{N_{i}}\frac{\s(v^{(i)}_j+v^{(i-1)}_k+\frac{w}{2})
 \s(v^{(i)}_j-v^{(i-1)}_k-\frac{w}{2})}{\s(v^{(i)}_j+v^{(i-1)}_k+\frac{3w}{2})
 \s(v^{(i)}_j-v^{(i-1)}_k+\frac{w}{2})}\no\\
 &&\qquad\qquad\qquad\qquad
 \times\prod_{k=1}^{N_{i+2}}\frac{\s(v^{(i)}_j+v^{(i+1)}_k+\frac{w}{2})
 \s(v^{(i)}_j-v^{(i+1)}_k-\frac{w}{2})}{\s(v^{(i)}_j+v^{(i+1)}_k+\frac{3w}{2})
 \s(v^{(i)}_j-v^{(i+1)}_k+\frac{w}{2})},\no\\
&&\hspace{22mm}i=0,\cdots,n-3,\label{BA1-1}\\
 &&
 \hspace{-24pt}
 \b_s^{(n-1)}(v^{(n-2)}_j)\frac{\s(2v^{(n-2)}_j\hspace{-0.08truecm}
 +\hspace{-0.08truecm}w)}{\s(2v^{(n-2)}_j\hspace{-0.08truecm}
 +\hspace{-0.08truecm}2w)}
 \prod_{k\neq j,k=1}^{N_{n-1}}\frac{\s(v^{(n-2)}_j+v^{(n-2)}_k)
 \s(v^{(n-2)}_j-v^{(n-2)}_k-w)}
 {\s(v^{(n-2)}_j\hspace{-0.08truecm}+\hspace{-0.08truecm}v^{(n-2)}_k
 \hspace{-0.08truecm}+\hspace{-0.08truecm}2w)
 \s(v^{(n-2)}_j\hspace{-0.08truecm}-\hspace{-0.08truecm}v^{(n-2)}_k
 \hspace{-0.08truecm}+\hspace{-0.08truecm}w)}\no\\
 &&~~~~=\frac{\s(\l_{s-1}+\bar{\xi}+v^{(n-2)}_j+w)
 \s(\l_{s-1}+\xi^{(n-1)}-v^{(n-2)}_j-\frac{w}{2})}
 {\s(\l_{s-1}+\bar{\xi}-v^{(n-2)}_j-w)\s(\l_{s-1}+\xi^{(n-1)}+v^{(n-2)}_j+\frac{w}{2})}\no\\
&&\qquad\qquad\qquad\qquad
\times\prod_{k=1}^{N_{n-2}}\frac{\s(v^{(n-2)}_j+v^{(n-3)}_k+\frac{w}{2})
\s(v^{(n-2)}_j-v^{(n-3)}_k-\frac{w}{2})}{\s(v^{(n-2)}_j+v^{(n-3)}_k+\frac{3w}{2})
 \s(v^{(n-2)}_j-v^{(n-3)}_k+\frac{w}{2})}.\label{BA1-2}
\end{eqnarray}
Here we have used the convention: $v^{(-1)}_k=z_k,\,
k=1,\cdots,N$. The quasi-classical property (\ref{quasi1}) of
$R(u,m)$, (\ref{function-a-1}) and (\ref{function-b}) lead to the
following relations
\begin{eqnarray}
 \b_s^{(i+1)}(u,\bar{\xi},0)=1,\quad \frac{\partial}{\partial
 u}\b_s^{(i+1)}(u,\bar{\xi},0)=0,\quad i=0,\cdots,n-2.\label{functionb-1}
\end{eqnarray}
Noticing the restriction (\ref{Restriction-1}), one may  introduce
some functions $\{\g_s^{(i+1)}(u)\}$ associated with
$\{\b_s^{(i+1)}(u,\bar{\xi},w)\}$
\begin{eqnarray}
 \g_s^{(i+1)}(u)=\frac{\partial}{\partial
 w}\b_s^{(i+1)}(u,\bar{\xi},w)|_{w=0}+\d\,
 \frac{\partial}{\partial
 \bar{\xi}}\b_s^{(i+1)}(u,\bar{\xi},0)|_{\bar{\xi}=\xi}
 ,\quad i=0,\cdots,n-2.\label{fuctiong}
\end{eqnarray}
Using (\ref{trans-2}), we can extract $n$ sets of eigenvalues
$\{h^{(s)}_j|s=1,\ldots,n\}$ (resp. the corresponding Bethe Ansatz
equations) of the Gaudin operators $H_j$ (\ref{Ham}) from  the
expansion around $w=0$ for the first order of $w$ of the
eigenvalues (\ref{Eigenvalue1}) of the transfer matrix $\t(u=z_j)$
(resp. the Bethe Ansatz equations (\ref{BA1-1}) and (\ref{BA1-2})
). Finally, the eigenvalues of the $\Zb_n$ elliptic Gaudin
operators are
\begin{eqnarray}
h^{(s)}_j=\g_s^{(1)}(z_j)+\zeta(\l_s+\xi+z_j)-\sum_{k=1}^{N_1}
\lt\{\zeta(z_j+x_k)+\zeta(z_j-x_k)\rt\},\label{Eig-1}\end{eqnarray}
where $\zeta$-function is defined in (\ref{Z-function}). The
$\frac{n(n-1)l}{2}$ parameters
$\{x^{(i)}_k|~k=1,2,\cdots,N_{i+1},~i=0,1,\cdots,n-2\}$ (including
$x_k$ as $x_k=x^{(0)}_k,\,k=1,\cdots, N_1$) are determined by the
following Bethe Ansatz equations
\begin{eqnarray}
 &&\g_s^{(i+1)}(x^{(i)}_j)-\zeta(2x^{(i)}_j)-2\sum_{k\neq
 j,k=1}^{N_{i+1}}\lt\{\zeta(x^{(i)}_j+x^{(i)}_k)+
 \zeta(x^{(i)}_j-x^{(i)}_k)\rt\}\no\\
 &&\qquad\qquad\quad =\g_s^{(i+2)}(x^{(i)}_j)-\sum_{k=1}^{N_{i}}
 \lt\{\zeta(x^{(i)}_j+x^{(i-1)}_k)+
 \zeta(x^{(i)}_j-x^{(i-1)}_k)\rt\}\no\\
 &&\qquad\qquad\qquad\qquad +\zeta(\l_{i+s+1}
 \hspace{-0.08truecm}+\hspace{-0.08truecm}\xi\hspace{-0.08truecm}+
 \hspace{-0.08truecm}x^{(i)}_j)
   \hspace{-0.08truecm}-\hspace{-0.08truecm}\sum_{k=1}^{N_{i+2}}
 \lt\{\zeta(x^{(i)}_j\hspace{-0.08truecm}+\hspace{-0.08truecm}x^{(i+1)}_k)
 \hspace{-0.08truecm}+\hspace{-0.08truecm}
 \zeta(x^{(i)}_j\hspace{-0.08truecm}-
 \hspace{-0.08truecm}x^{(i+1)}_k)\rt\},\no\\
&&\hspace{27mm}i=0,\cdots,n-3,\label{BAE-1}\\
 &&\g_s^{(n-1)}(x^{(n-2)}_j)
\hspace{-0.08truecm}
-\hspace{-0.08truecm}\zeta(2x^{(n-2)}_j)\hspace{-0.08truecm}-
\hspace{-0.08truecm}2\hspace{-0.08truecm}\sum_{k\neq
 j,k=1}^{N_{n-1}}\lt\{\zeta(x^{(n-2)}_j\hspace{-0.08truecm}+
 \hspace{-0.08truecm}x^{(n-2)}_k)\hspace{-0.08truecm}+\hspace{-0.08truecm}
 \zeta(x^{(n-2)}_j\hspace{-0.08truecm}-\hspace{-0.08truecm}x^{(n-2)}_k)\rt\}\no\\
 &&\qquad\qquad\quad =\lt(\d+\frac{n}{2}\rt)\zeta(\l_{s-1}\hspace{-0.08truecm}+
 \hspace{-0.08truecm}\xi\hspace{-0.08truecm}+\hspace{-0.08truecm}x^{(n-2)}_j)
 \hspace{-0.08truecm}+\hspace{-0.08truecm}
 \lt(\frac{2-n}{2}-\d\rt)\zeta(\l_{s-1}\hspace{-0.08truecm}+\hspace{-0.08truecm}
 \xi\hspace{-0.08truecm}-\hspace{-0.08truecm}x^{(n-2)}_j)\no\\
 &&\qquad\qquad\qquad\qquad\qquad\quad\quad -\sum_{k=1}^{N_{n-2}}
 \lt\{\zeta(x^{(n-2)}_j\hspace{-0.08truecm}+\hspace{-0.08truecm}x^{(n-3)}_k)
 \hspace{-0.08truecm}+\hspace{-0.08truecm}
 \zeta(x^{(n-2)}_j\hspace{-0.08truecm}-\hspace{-0.08truecm}
 x^{(n-3)}_k)\rt\}.\label{BAE-2}
\end{eqnarray}Here we
have used the convention: $x^{(-1)}_k=z_k,\,k=1,\cdots,N$ in
(\ref{BAE-1}). Then the $n$ sets of eigenvalues
$\{h^{(s)}_j|s=1,\ldots,n\}$ given by (\ref{Eig-1})-(\ref{BAE-2})
(c.f. \cite{Yan04-4}) together constitute the complete spectrum of
the Gaudin operators $H_j$ (\ref{Ham}).

\section{Conclusions} \label{Con}
\setcounter{equation}{0}

We have discovered the multiple reference state structure of the
$\Zb_n$ Belavin model with boundaries specified by the
non-diagonal K-matrices $K^{\pm}(u)$, (\ref{K-matrix}) and
(\ref{K-matrix1}). It is found that there exist $n$ reference
states $\{|\O^{(s)}(\l)\rangle|s=1,\ldots,n\}$ (\ref{Vac}), which
lead to $n$ sets of Bethe states
$|v_1,\cdots,v_{N_1}\rangle^{(s)}$ (\ref{Eigenstate}). These Bethe
states give rise to $n$ sets of Bethe Ansatz equations
(\ref{BA1})-(\ref{BA2}) and eigenvalues (\ref{Eigenvalue1}),
labelled by $s=1,\ldots,n$. The fist set of them, which
corresponds to the $s=1$ case, gives the results found in
\cite{Yan04}. It is expected that these $n$ sets of eigenvalues
$\{\L_s(u;\xi,\{v_{k}\})|s=1,\ldots,n \}$ (\ref{Eigenvalue1})
together give rise to the complete spectrum of the transfer matrix
$\t(u)$ (\ref{trans}) for the $\Zb_n$ Belavin model with generic
boundaries. In the quasi-classical limit (i.e. $w\rightarrow 0$),
the resulting $n$ sets of eigenvalues $\{h^{(s)}_j|s=1,\ldots,n\}$
given by (\ref{Eig-1})-(\ref{BAE-2}) together constitute the
complete spectrum of the Gaudin operators $H_j$ (\ref{Ham}).

Taking the scaling limit \cite{Yan01} of our general results, for
the special $n=2$ case, we recover the results obtained in
\cite{Yan07} for the open XXZ spin chain. It is believed  that
such structure of multiple reference states also exists for the
open  $A^{(1)}_{n-1}$ trigonometric vertex model studied in
\cite{Yan05}.

\section*{Acknowledgements}
The financial support from  Australian Research Council  is
gratefully acknowledged.


\section*{Appendix A: Definitions of the $\a$-, $\b$-, $\L$- functions}
\setcounter{equation}{0}
\renewcommand{\theequation}{A.\arabic{equation}}

In this appendix, we give the definitions of the functions
$\a^{(i)}_s(u)$, $\b^{(i)}_s(u)$ and
$\L^{(i)}(u;\xi,\{v^{(i)}_{k}\})$, which appeared in the
expressions of the eigenvalues and the Bethe Ansatz equations
(\ref{BA1})-(\ref{Eigenvalue1}).

In order to carry out the nested Bethe Ansatz for the $\Zb_n$
Belavin model with the generic open boundary conditions, one needs
to introduce a set of reduced K-matrices $\tilde{\K}^{(r)}(\l|u)$
which include the original one
$\tilde{\K}(\l|u)=\tilde{\K}^{(0)}(\l|u)$:
\begin{eqnarray}
 \tilde{\K}^{(r)}(\l|u)^j_i&=&
 \d^j_i\,\tilde{k}^{(r)}(\l|u)_i,\quad i,j=r+1,\cdots,n,
 \quad r=0,\cdots,n-1,\no\\
 \tilde{k}^{(r)}(\l|u)_i&=& \lt\{\prod_{k\neq
 i,k=r+1}^n\frac{\s(\l_{ik}-w)}{\s(\l_{ik})}\rt\}
 \frac{\s(\l_i+\bar{\xi}+u+
 \frac{(n-r)w}{2})}{\s(\l_i+\bar{\xi}-u-\frac{(n-r)w}{2})}.\no
\end{eqnarray}
Moreover we introduce a set of functions
$\{\a^{(r)}(u)|r=1,\cdots,n-1\}$  related to the reduced
K-matrices $\tilde{\K}^{(r)}(\l|u)$
\begin{eqnarray}
 \a^{(r)}(u)=\sum_{i=r}^{n}R(2u,\l+w\hat{r})^{ir}_{ri}\,
 \tilde{k}^{(r-1)}(\l|u)_i,\quad r=1,\cdots,n-1,\label{function-a-1}
\end{eqnarray} and an associated set of functions
$\{\b^{(i)}(u,\bar{\xi},w)|i=1,\ldots, n-1\}$
\begin{eqnarray}
 \b^{(i+1)}(u,\bar{\xi},w)\equiv
 \b^{(i+1)}(u)=\a^{(i+1)}(u)\frac{\s(\l_{i+1}+\xi-u-\frac{i}{2}w)}
 {\s(\l_{i+1}+\xi+u+w-\frac{i}{2}w)},~~i=0,\cdots,
 n-2.\label{function-b}
\end{eqnarray} In the process of carrying out the nested Bethe
Ansatz \cite{Yan04}, one needs to introduce a set of functions
$\{\L^{(i)}(u;\xi,\{v^{(i)}_{k}\})|i=0,\ldots,n-1\}$ which
correspond to the eigenvalues of the reduced transfer matrices.
The functions $\{\L^{(i)}(u;\xi,\{v^{(i)}_{k}\})\}$ are given by
the following recurrence relations
\begin{eqnarray}
 \L^{(i)}(u;\xi^{(i)},\{v^{(i)}_k\})
 \hspace{-0.18truecm}&=&\hspace{-0.18truecm}\b^{(i+1)}(u)
 \frac{\s(\l_{i+1}+\xi^{(i)}+u+w)}{\s(\l_{i+1}+\xi^{(i)}+u)}
 \prod_{k=1}^{N_{i+1}}\frac{\s(u+v^{(i)}_k)\s(u-v^{(i)}_k-w)}
 {\s(u+v^{(i)}_k+w)\s(u-v^{(i)}_k)}\no\\
 \hspace{-0.18truecm}&&\hspace{-0.18truecm}
 +\frac{\s(2u)\s(\l_{i+1}\hspace{-0.08truecm}+\hspace{-0.08truecm}
 u\hspace{-0.08truecm}+\hspace{-0.08truecm}w\hspace{-0.08truecm}
 +\hspace{-0.08truecm}\xi^{(i)})}
 {\s(2u\hspace{-0.08truecm}+\hspace{-0.08truecm}w)\s(\l_{i+1}
 \hspace{-0.08truecm}+\hspace{-0.08truecm}u\hspace{-0.08truecm}
 +\hspace{-0.08truecm}\xi^{(i)})}
 \prod_{k=1}^{N_{i+1}}\frac{\s(u\hspace{-0.08truecm}-\hspace{-0.08truecm}
 v^{(i)}_k\hspace{-0.08truecm}+\hspace{-0.08truecm}w)
 \s(u\hspace{-0.08truecm}+\hspace{-0.08truecm}v^{(i)}_k
 \hspace{-0.08truecm}+\hspace{-0.08truecm}2w)}
 {\s(u\hspace{-0.08truecm}-\hspace{-0.08truecm}v^{(i)}_k)
 \s(u\hspace{-0.08truecm}+\hspace{-0.08truecm}v^{(i)}_k
 \hspace{-0.08truecm}+\hspace{-0.08truecm}w)}\no\\
 \hspace{-0.18truecm}&&\hspace{-0.18truecm}\,\times
 \hspace{-0.08truecm}
 \prod_{k=1}^{N_i}\frac{\s(u+z^{(i)}_k)\s(u-z^{(i)}_k)}
 {\s(u\hspace{-0.08truecm}+\hspace{-0.08truecm}
 z^{(i)}_k\hspace{-0.08truecm}+\hspace{-0.08truecm}w)
 \s(u\hspace{-0.08truecm}-\hspace{-0.08truecm}z^{(i)}_k
 \hspace{-0.08truecm}+\hspace{-0.08truecm}w)}
 \L^{(i+1)}(u\hspace{-0.08truecm}+\hspace{-0.08truecm}
 \frac{w}{2};\xi^{(i)}\hspace{-0.08truecm}-
 \hspace{-0.08truecm}\!\frac{w}{2},\{v_j^{(i+1)}\}),\no\\
 \hspace{-0.18truecm}&&\hspace{-0.18truecm}
 ~~~~~~~~~~~~i=1,\cdots,n-2,\label{Eigenvalue2} \\
 \L^{(n-1)}(u;\xi^{(n-1)})&=&\frac{\s(\l_n+\bar{\xi}+u+\frac{w}{2})
 \s(\l_n+\xi^{(n-1)}-u)} {\s(\l_n+\bar{\xi}-u-\frac{w}{2})
 \s(\l_n+\xi^{(n-1)}+u)}.\label{Eigenvalue3}
\end{eqnarray}
The reduced boundary parameters $\{\xi^{(i)}\}$ and inhomogeneous
parameters $\{z^{(i)}_k\}$ are given by (\ref{parameters}). It is
remarked that all the functions $\{\a^{(i)}(u)\}$,
$\{\b^{(i)}(u)\}$ and $\{\L^{(i)}(u;\xi,\{v^{(i)}_{k}\})\}$ are
{\it indeed} the functions of the boundary parameters $\{\l_i\}$.

Since that the suffix index of the boundary parameters $\{\l_i\}$
takes value in $\Zb_n$, one can introduce a $\Zb_n$ cyclic
operator $\P$ (i.e. $\P^n={\rm id}$), which acts on the space of
functions of $\{\l_1,\ldots,\l_n\}$. On any function
$f(\l_1,\ldots,\l_n)$ the action of the operator $\P$ is given by
\begin{eqnarray}
 \P\,(f(\l_1,\ldots,\l_{n-1},\l_n))
 =f(\l_{1+1},\ldots,\l_{n},\l_{n+1})
 =f(\l_{2},\ldots,\l_{n},\l_{1}).\label{Action}
\end{eqnarray}

Then we introduce the following $\a$-, $\b$-,$\L$-functions:
\begin{eqnarray}
 \a^{(i)}_s(u)&=&\P^{s-1}\,(\a^{(i)}(u)),\quad s=1,\ldots
 n,\quad i=1,\ldots n-1,\label{alpha-functions}\\
 \b^{(i)}_s(u)&=&\P^{s-1}\,(\b^{(i)}(u)),\quad s=1,\ldots
  n,\quad i=1,\ldots n-1,\label{beta-functions}\\
 \L^{(i)}_s(u;\xi,\{v^{(i)}_{k}\})&=&\P^{s-1}\,
 \L^{(i)}(u;\xi,\{v^{(i)}_{k}\}),\quad s=1,\ldots
 n,\quad i=1,\ldots n-1.\label{lemada-functions}
\end{eqnarray}Direct calculation shows that the two definitions of
$\a^{(1)}_s$, (\ref{function-a}) and (\ref{alpha-functions}),
coincide with each other.


\end{document}